\documentstyle[12pt,aaspp4]{article}
\def\clock{\count0=\time \divide\count0 by 60
     \count1=\count0 \multiply\count1 by -60 \advance\count1 by \time
     \number\count0:\ifnum\count1<10{0\number\count1}\else\number\count1\fi}

\begin{document}
\title{Normalizing the Temperature Function of Clusters of Galaxies}
\author{Ue-Li Pen\altaffilmark{1}}
\affil{Harvard Society of Fellows and Harvard-Smithsonian Center
for Astrophysics}
\altaffiltext{1}{e-mail I: upen@cfa.harvard.edu}
\newcommand{\rg}{\sqrt{g}}
\newcommand{\Nabla}{\bigtriangledown}
\newcommand{\etal}{{\it et al.\ }}

\begin{abstract}
We re-examine the constraints which can be robustly obtained from the
observed temperature function of X-ray cluster of galaxies.  
The cluster mass function has
been thoroughly studied in simulations and analytically, but a direct
simulation of the temperature function is presented here for the first
time.  Adaptive hydrodynamic simulations using the cosmological Moving
Mesh Hydro code of Pen (1997a) are used to calibrate the temperature
function for different popular cosmologies.  Applying the new
normalizations to the present-day cluster abundances, we find
$\sigma_8=0.53\pm 0.05 \Omega_0^{-0.45}$ for a hyperbolic universe, and
$\sigma_8=0.53\pm 0.05 \Omega_0^{-0.53}$ for a spatially flat universe
with a cosmological constant.  The simulations followed the
gravitational shock heating of the gas and dark matter, and used a
crude model for potential energy injection by supernova heating.  The
error bars are dominated by uncertainties in the heating/cooling
models.  We present fitting formulae for the mass-temperature
conversions and cluster abundances based on these simulations.
\end{abstract}

\section{Introduction}

X-ray studies of clusters of galaxies have provided a host of
quantitative data for the study of cosmology.  They are ideally suited
for the normalization of structure formation since their mass
$(\approx 10^{15} $M$_{\odot})$ is close to the mass formed from the
collapse of an $8h^{-1}$ Mpc radius sphere of matter.  They provide a
direct dynamical measure $\sigma_8$, the linearly extrapolated
variation in mass in such spheres.  We can numerically
compute their formation directly through simulations from linear
perturbation theory into the non-linear regime with appropriate
boundary conditions.   Only about a percent of the mass of the
universe is in virialized rich clusters.   For a Gaussian
random field the variance depends only logarithmically on the
abundance, making an accurate measurement possible even in the
presence of substantial Poissonian errors.

While the gravitational mass in bound objects is easily computed using
analytic approaches such as the Press-Schechter formalism (Press and
Schechter 1974) (hereafter P-S) or N-body simulations, it is not so
easily measured.  The most robust measurements of the cluster
abundance currently rely on the cluster temperature function, the
number density of clusters above some temperature $N(>kT)$ expressed
in units of $h^3$ Mpc$^{-3}$.  Henry and Arnaud (1991) (hereafter
referred to as HA) and Edge \etal (1990) have analysed the HEAO-1 A2
all sky X-ray survey to obtain temperatures for clusters at a flux
limit of $F_{2-10\rm keV} > 3\times 10^{-11}$ erg/cm$^2$/sec.  The
sample contains the 25 X-ray brightest clusters and is over 90\%
complete at galactic latitude $|b|>20^o$.  By virtue of their
brightness the temperatures are easily and well determined.  In this
paper we will make use of this excellent objective sample.

From the theoretical stand point, the cluster temperature function is
a clean measurement since the normalization of fluctuations can be
established independently of the baryon fraction $\Omega_b$, or the
Hubble constant $H_0\equiv 100 h$km/sec/Mpc.  Clusters are close to
isothermal, both observationally and in simulations, which makes their
temperature determination robust and insensitive to numerical
resolution or  telescope angular resolution.  The temperature of a
cluster depends primarily on the depth of the potential well of the
dark matter, and the state of equilibrium of the gas.  This is in
contrast to the cluster luminosity, which depends strongly on small
scale parameters like clumping and core radius.  While heating or
cooling processes could change the temperatures slightly, the
gravitational potential provided by the dark matter buffers the system
and causes the temperatures in hydrostatic equilibrium to only be
weakly affected by energy injection or loss.  We will verify this
effect through direct simulations of heat injection.  Furthermore,
measurements of the galaxy velocity dispersion appears to be in good
agreement with the temperatures, with the ratio of gas temperature to
velocity dispersion $\beta_{\rm fit} \equiv kT/\mu m_p \sigma^2 =
0.95\pm 0.05$ (Bahcall and Lubin 1994).  This suggests that
non-thermal pressures are negligible.

Recent work by Eke, Cole and Frenk (1995) (hereafter ECF), and by
Viana and Liddle (1996) (hereafter VL) used the cluster temperature
function to normalise the amplitude of fluctuations in spheres of
radius $8 h^{-1}$ Mpc, denoted $\sigma_8$.  They compared the observed
cluster temperature function of HA to Press-Schechter calculations
calibrated by N-body simulations.  These studies did not include any
gasdynamic effects, leaving open questions about the statistical
properties of cluster equilibria.  They concluded that the present day
normalization in a flat standard CDM is $\sigma_8=0.5$.  Modelling the
predicted cluster abundance in a subcritical density universe is more
difficult.  VL propose $\sigma_8 \propto \Omega_0^{-0.5\sim -0.7}$,
while ECF have a smaller error budget in the exponent.  ECF only
measured the mass function in simulations.  In order to convert from a
mass function to a temperature function, these models all assumed that
all clusters are perfect singular isothermal spheres which formed at
some prescribed redshift.  While the errors in the normalization
$\sigma_8$ from the Poissonian scatter are very small, any error in
mass-temperature conversion translates into a similar sized error in
the normalization.  That issue is the point of largest uncertainty in
these semi-analytic calculations.  In order to address this problem,
we extend their work by performing full hydrodynamic simulations. We
directly measure the X-ray emission weighted temperature in the
simulation.  The simulations are used to directly measure the
mass-temperature relation which are then fed into the P-S
calculations.  This approach allows us to explore a large dynamic
range in volumes and parameters.

In this paper we will not use the term ``open universe'', but instead
call universes with negative spatial curvature ``hyperbolic
universe''.  This circumvents the misleading suggestion that
hyperbolic universes should be spatially infinite (Pen and Spergel
1995).  We will proceed in section \ref{sec:sim} to describe the
simulations.  In section \ref{sec:otemp} we summarize the observed
cluster temperature function.  Our new results are presented in
section \ref{sec:mtr}.  We discuss the cosmological implication of
these results in section \ref{sec:cosmo}.

\section{Simulations}
\label{sec:sim}
We use the Moving Mesh Hydrodynamic (MMH) code (Pen 1997a).  It
implements a Total Variation Diminishing (TVD) (Xin and Jin 1995) high
resolution shock capturing hydrodynamics scheme on an adaptively
deforming mesh.  The gravitational potential is solved using a
multigrid iteration (Pen 1995).  Dark matter is modeled using a
particle-mesh algorithm on the same moving grid.  The grid is
continuously adjusted to maintain an approximately constant mass per
cell.  This is achieved through a pure potential flow grid velocity
field.  The full Euler fluid equations are solved on this moving mesh,
and the fluid is allowed to develop vorticities.  By following the
mass, the MMH code has improved spatial resolution in dense regions
such as clusters of galaxies.  Since their cores are $10^3-10^4$ times
overdense, any mass based method will have a ten to twenty fold better
length resolution in these high density regions.  A recent study has
compared this code to several other cosmological hydrodynamic codes
for a cluster formation scenario (Frenk \etal\ 1997), and good general
agreement exists between this code and the others compared.  Excellent
agreement was found for the temperature properties of simulated
clusters.  Similar agreement was obtained in a comparison to a suite
of existing codes (Pen 1997a, Kang \etal\ 1994).  All simulations were
run on an SGI power challenge at the National Center for
Supercomputing Applications.  They all used $128^3$ grid cells and
$256^3$ particles.  The initial power spectrum was taken from Bardeen
\etal (1986) (hereafter called BBKS).  The simulations were started at
a redshift $z=100$.  Table \ref{tbl:sim} summarizes the cosmological
parameters used in each simulation.  Cooling has not been incorporated
in these simulations, and will be addressed in a future paper (Cen
\etal 1998).

\begin{table}
\begin{center}
\begin{tabular}{|l|l|l|l|l|l|}
\hline
model 	& $\Omega_0$ 	&$\Omega_\Lambda$& $\sigma_8$ 	&$h$&L($h^{-1}$ Mpc) \\
\hline\hline
CDM1-3 	&	1	&	0	&	0.5	& 0.5	&80 \\
CDM4 	&	1	&	0	&	0.5	& 0.5	&200 \\
PREHEAT	&	1	&	0	&	0.5	& 0.5	&80 \\
OCDM	&	0.37	&	0	&	1	& 0.7	&120 \\
LCDM	&	0.37	&	0.63	&	1	& 0.7	&120\\
\hline
\end{tabular}
\end{center}
\caption{Simulations used for this study.  All simulations used a
$128^3$ grid with $256^3$ particles.  All models have a baryonic
fraction $\Omega_b h^2 =0.0125$ and compression limiter $c_{\rm
max}=10$.  The effective resolution is $(128 c_{\rm max})^3$ (Pen
1997a). The two low $\Omega$ models used identical random seeds.}  
\label{tbl:sim}
\end{table}

For the CDM model we used the best fit values suggested by ECF with
$\Omega_0=1,\ \Omega_b=0.05,\ h=0.5, \ \sigma_8=0.5, \ n=1$ on a box
of side length $80 h^{-1}$ Mpc.  $n$ denotes the unprocessed power
spectrum (BBKS), whose Harrison-Zeldovich-Peebles (hereafter HZP)
value is $n=1$.  The simulations were repeated two
times with different random seeds to improve the statistical measures.
These comprise models CDM1-3.  We note that the finite box sizes
truncates the long wave modes, which has the effect of lowering the
fluctuation amplitude in the numerical realization by 15\% (Gelb and
Bertschinger 1994, Pen 1997b).  To check its significance, and to test
the resolution dependence, we ran a simulation in a larger box of side
length $200 h^{-1}$ Mpc which we call CDM4.  At such small values of
$\Omega_b$ the corrections to the power spectrum are negligible
compared to other sources of error (Holtzman 1989).  As long as the
gravitational potential is dominated by the dark matter, we can
rescale the result to any value of $\Omega_b$ easily.  We then ran a
model with heat input from early star formation, which we call
PREHEAT.  At a redshift $z=1$, we injected 1 keV of energy per nucleon
into the plasma.  Details are described in section \ref{sec:mtr}.  We
next ran a cosmological constant model $\Omega_\Lambda=0.63, \
\sigma_8=1$ in a box of side length $120 h^{-1}$ Mpc.  We repeated the
simulation with identical initial conditions for a hyperbolic universe
$\Omega_0=0.37$.  These are called LCDM and OCDM respectively.

Since clusters are rare objects, their identification is a relatively
simple matter.  We use the gas densities to identify clusters.  We
search the volume for all density peaks in the unsmoothed gas field
which are overdense by at least 200 over the mean cosmic density, and
separated by at least $2h^{-1}$ Mpc.  If two peaks are closer than
that distance, the less dense one is discarded.  From this sample, we
computed the total emission weighted temperature for each cluster in a
$1h^{-1}$ Mpc radius, weighting the temperature in each cell by
$\rho^2 T^{1/2}$, and sorted these clusters by temperature to find the
cumulative temperature function.

\section{Observed Temperature Function}
\label{sec:otemp}
\begin{figure}
\plotone{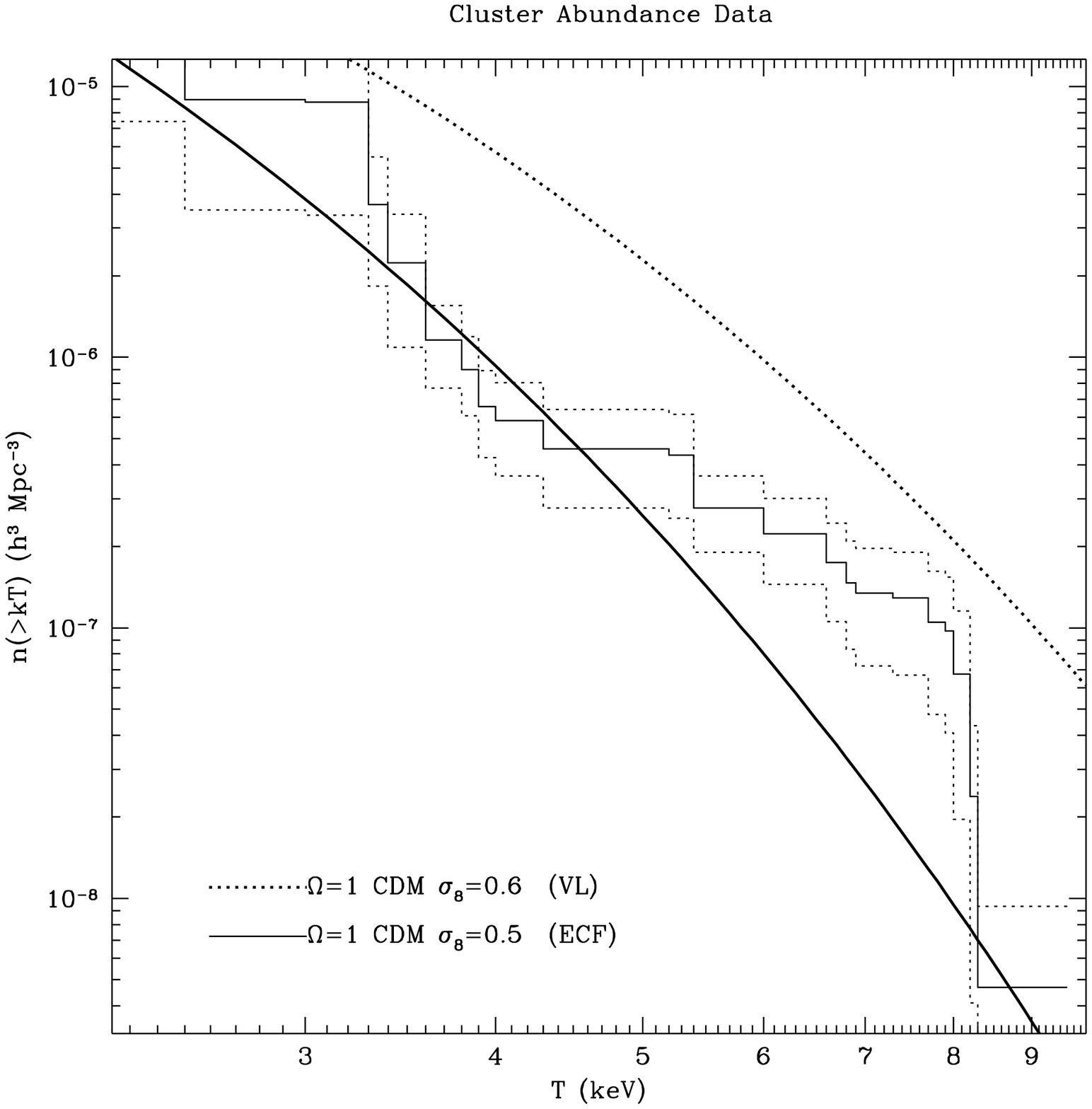}
\caption{The stepped solid line is the observed cumulative temperature
function from HA.  The upper and lower thin stepped lines 
indicate $1-\sigma$ Poisson errors.
The diagonal solid curve represents the cluster abundance normalized
to $\sigma_8=0.5$ (ECF), while the dotted curve is the VL
normalization $(\sigma_8=0.6)$.} 
\label{fig:hadata}
\end{figure}

We  use the cluster sample from HA.  The sample is based on the
objects from the HEAO-1 A2 survey, 
which identified all objects at a flux limit $F_x > 3\times 10^{-11}$
erg/cm$^2$/sec in the $2-10$ keV band at a galactic latitude
$|b|>20^o$.  The 2-10 keV window over which HEAO is sensitive is well
matched to the temperatures of rich clusters.
HA identified 25 clusters with that sample.  They range
in temperature from $2.5-9.5$ keV, with luminosities in the range
$10^{43}-10^{45}h^{-2}$ erg/sec.  The furthest cluster is at redshift
$z=0.09$, which is sufficiently low that evolutionary corrections are
small.  This sample has excellent completeness properties and most 
clusters have multiply measured temperatures.  The typical temperature
error is 10\%.
Following ECF, we measure the cumulative cluster density
\begin{equation}
N(>kT) = \sum_{kT_i>kT} \frac{1}{V_{\rm max,i}}
\label{eqn:vmax}
\end{equation}
where $V_{\rm max,i}$ is the maximal volume to which each cluster
could have been seen at the flux limit.
The formal Poisson error on the estimate is given by
\begin{equation}
\sigma(>kT)^2 = \sum_{kT_i>kT} \frac{1}{V^2_{\rm max,i}}.
\label{eqn:var}
\end{equation}
This tends to underestimate the true error for several reasons.  Since
we are measuring a cumulative abundance, errors at different
temperatures are correlated.  Clusters tend to cluster strongly with
each other with correlation lengths of about $20 h^{-1}$ Mpc (Bahcall
1996).  This boosts the error in (\ref{eqn:var}) for the low
luminosity clusters where $V_{\rm max}$ is comparable to the
correlation volume.  The errors
are Poissonian with respect to the true underlying distribution, not
with repect to the estimated density.  Using the estimated density
will again systematically underestimate the errors.
Instead of attempting to model these uncertain errors directly, we
will use a heuristic fit based on the estimates of ECF.
They found that the abundance
is best represented near 5 keV, which lies in the center of the
temperature window.  This value is close to the temperature formed by
the collapse of an $8h^{-1}$ Mpc radius sphere in a $\Omega=1$ model,
and empirically 
appears to be the pivot point of the distribution as one varies
parameters.  Normalizing abundances at that temperature, ECF found
statistical errors result in variations of $\sigma_8$ of only about
2\%.  As we will see below, the errors arising from thermal history
uncertainties and numerical limitations are significantly larger, so
we will neglect the Poissonian errors hereafter.
The cluster abundance and various past fits are shown in Figure
\ref{fig:hadata} renormalized using the results of this paper (see
equations \ref{eqn:asymptsimp}, \ref{eqn:mtrz} and \ref{eqn:mtr} below).

\section{Mass-Temperature Relation}
\label{sec:mtr}

\begin{figure}
\plotone{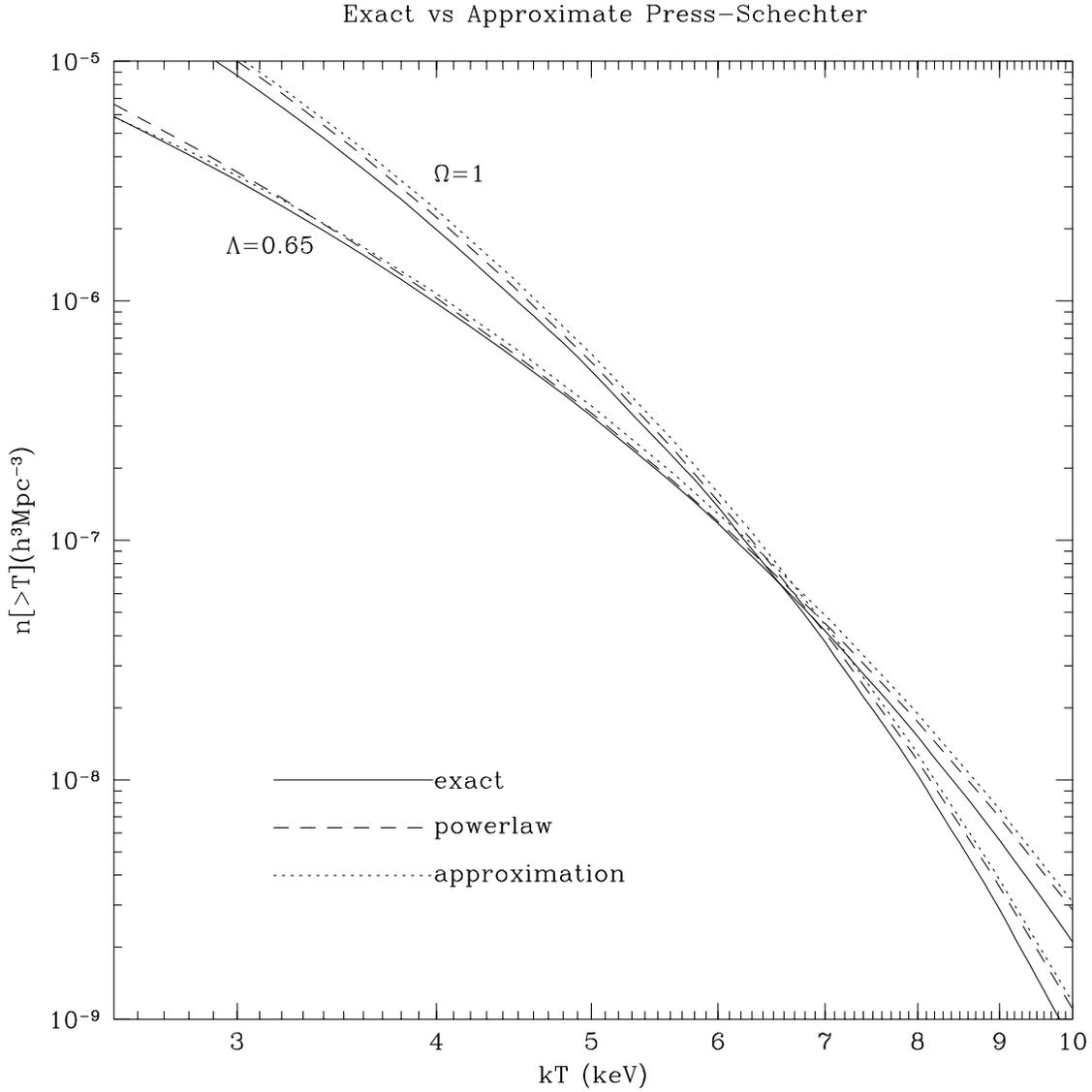}
\caption{Comparison between the temperature function obtained by
integrating a full power spectrum compared to the scale-free power
law and its approximation.  We see that Equation
\protect{\ref{eqn:asymptsimp}} a good 
approximation on the scales of interest.}
\label{fig:compps}
\end{figure}

To obtain an analytic estimate of the temperature function, we will
use the Press-Schechter {\it Ansatz}.  The fraction of mass
in bound objects is
\begin{equation}
f(>M) = \sqrt{\frac{2}{\pi}} \int_\frac{\delta_c}{\sigma(M)}^\infty
e^{-u^2/2} du.
\label{eqn:pscum}
\end{equation}
$\sigma(M)$ is defined as the RMS density fluctuations in tophat
spheres of mass $M$.  The distribution function has been multiplied by
2 such that all the mass is accounted for when $\sigma \rightarrow
\infty$. 
$\delta_c = 1.686$ is the linearly extrapolated
overdensity at which an object virializes.  Formally it is only exact
for $\Omega_0=1$, but it varies only by a few percent to $\Omega_0=0.3$ and
we will consider it to be constant.    We will
numerically 
normalize the mass-temperature relation, which will absorb any changes
in $\delta_c$ and simplifies calculations.
We define the
dimensionless mass $m \equiv M/M_8$
where $M_8 \equiv 4\pi \bar{\rho}
(8 h^{-1} {\rm Mpc})^3/3$ is the mass contained in an $8h^{-1}$Mpc
sphere.  $\bar{\rho} \equiv 3\Omega_0 H_0^2/8\pi G$ is the mean density
of the universe today in terms of the Hubble constant $H_0 = 100
h$km/sec/Mpc.
Differentiating (\ref{eqn:pscum}) we obtain the differential number
density of objects $dn/dm=(\bar{\rho}/M_8 m) df/dm$ as
\begin{equation}
\frac{dn}{dm} = \sqrt{\frac{2}{\pi}} \frac{\bar{\rho}}{M_8}
\frac{\delta_c}{\sigma m^2} \left| \frac{d \ln \sigma}{d \ln m}
\right| \exp(\frac{-\delta_c^2}{2 \sigma^2}).
\label{eqn:diffps}
\end{equation}
The observed abundance (\ref{eqn:vmax}) is a cumulative statistic, so
it is desirable to 
integrate the differential density function (\ref{eqn:diffps}).
To simplify the algebra, we will assume a
pure powerlaw dependence $\sigma = \sigma_8 m^{-\alpha}$.  $\alpha$ is
related to the effective power spectrum $P(k)=k^{-3\alpha}$.
For CDM-like power spectra, the BBKS fit to the power spectrum 
at $r=8 h^{-1}$ Mpc has
\begin{equation}
\alpha=0.222+0.2495\Gamma - 0.0232\Omega_0,
\label{eqn:alpha}
\end{equation}
where 
$\Gamma \equiv \Omega_0 h
\exp[-\Omega_b -(\Omega_b/\Omega_0)]$ (VL).  (\ref{eqn:alpha}) is defined
by requiring the ratio of $\sigma_8/\sigma_X$ to be exact, where $X$ is
the radius which forms a 5 keV cluster in the three cosmological
models which were simulated.
We can then integrate
(\ref{eqn:diffps})
\begin{equation}
n(>m) = \frac{3}{2048 \pi^{3/2}} \left(
\frac{\delta_c^2}{2\sigma_8^2} \right)^{\frac{1}{2\alpha}}
\Gamma[\frac{\alpha-1}{2\alpha},
m^{2\alpha}\frac{\delta_c^2}{2\sigma_8^2}] \ h^{3} {\rm Mpc}^{-3}
\label{eqn:cumdist}
\end{equation}
where $\Gamma[a,x]$ is the incomplete $\Gamma$ function.  The
cumulative number abundance (\ref{eqn:cumdist}) has units of clusters per
Mpc$^3/h^3$.  The
asymptotic semi-convergent sequence for $\Gamma$ (Arfken 1985) allows
us to expand (\ref{eqn:cumdist})
\begin{equation}
n(>m) \simeq 2.63 \times 10^{-4} (\mu^{1/2\alpha}) \frac{\exp(-v)}{v^p}
\left(1-\frac{p}{v}+\frac{p(p+1)}{v^2}-\frac{p(p+1)(p+2)}{v^3} +
\cdots \right) \  h^{3} {\rm Mpc}^{-3}
\label{eqn:asympt}
\end{equation}
where we have abbreviated $\mu=\delta_c^2/2\sigma_8^2$, $v=\mu
m^{2\alpha}$, $p=(\alpha+1)/2\alpha$.  The terms in (\ref{eqn:asympt})
initially converge for large $v$, but at some point diverge again.
Each term has an alternating sign, and each approximation brackets the
true solution.  For our purposes, we can truncate the sum and modify
the first two terms as
\begin{equation}
n(>m) \simeq 2.63 \times 10^{-4} \mu^{1/2\alpha} \frac{\exp(-v)}{v^p}
\left(1-\frac{p}{2v} \right) h^{3} {\rm Mpc}^{-3}.
\label{eqn:asymptsimp}
\end{equation}
The accuracy of (\ref{eqn:asymptsimp}) is shown in Figure \ref{fig:compps},
where we have used the mass-temperature relation derived below.  Three
sets of lines are shown, corresponding to an integral of (\ref{eqn:diffps})
using the BBKS power spectrum (solid lines), the abundance using the
power law fit with (\ref{eqn:alpha}) and (\ref{eqn:cumdist}) (dashed
lines) and the two term asymptotic expansion (\ref{eqn:asympt})
(dotted lines).  For this illustration, we picked one model with
$\Omega=1$, and one cosmological constant model with
$\Omega_\Lambda=0.65$.  We see that the approximations are accurate to
about 3\% in temperature over the temperature range 3-10 keV.

For the next step, we need to convert the cumulative mass function
into a temperature function.  Various approaches, for example ECF,
assumed all objects to be perfect isothermal spheres forming at
$z=0$.  In the chain of calculations, this step is the least
certain link, and it is the purpose this paper to quantify it.
We will continue  
to rely on the virial relation $m \propto T^{3/2}$, which follows if
the formation redshift and state of equilibrium is uncorrelated with
cluster mass.  Bryan and Norman (1997) recently demonstrated this
relationship to
hold to a good approximation in hydrodynamic simulations.  We then
only need to know the  relation coefficient $T_8$
\begin{equation}
m = \left(\frac{kT}{T_8(1+z)}\right)^{3/2}
\label{eqn:mtrz}
\end{equation}
where $T_8$ is the effective temperature of a cluster which forms during the
collapse of an $8h^{-1}$ Mpc radius sphere.  
In a low density universe the same sphere contains much less mass, and
consequently $T_8$ depends on $\Omega_0$.
While the overdensity parameter $\delta_c$ in the top hat model has a
weak $\Omega_0$ dependence, we 
will fix its value at $\delta_c=1.686$ and absorb all modeling into
$T_8$.

\begin{figure}
\plotone{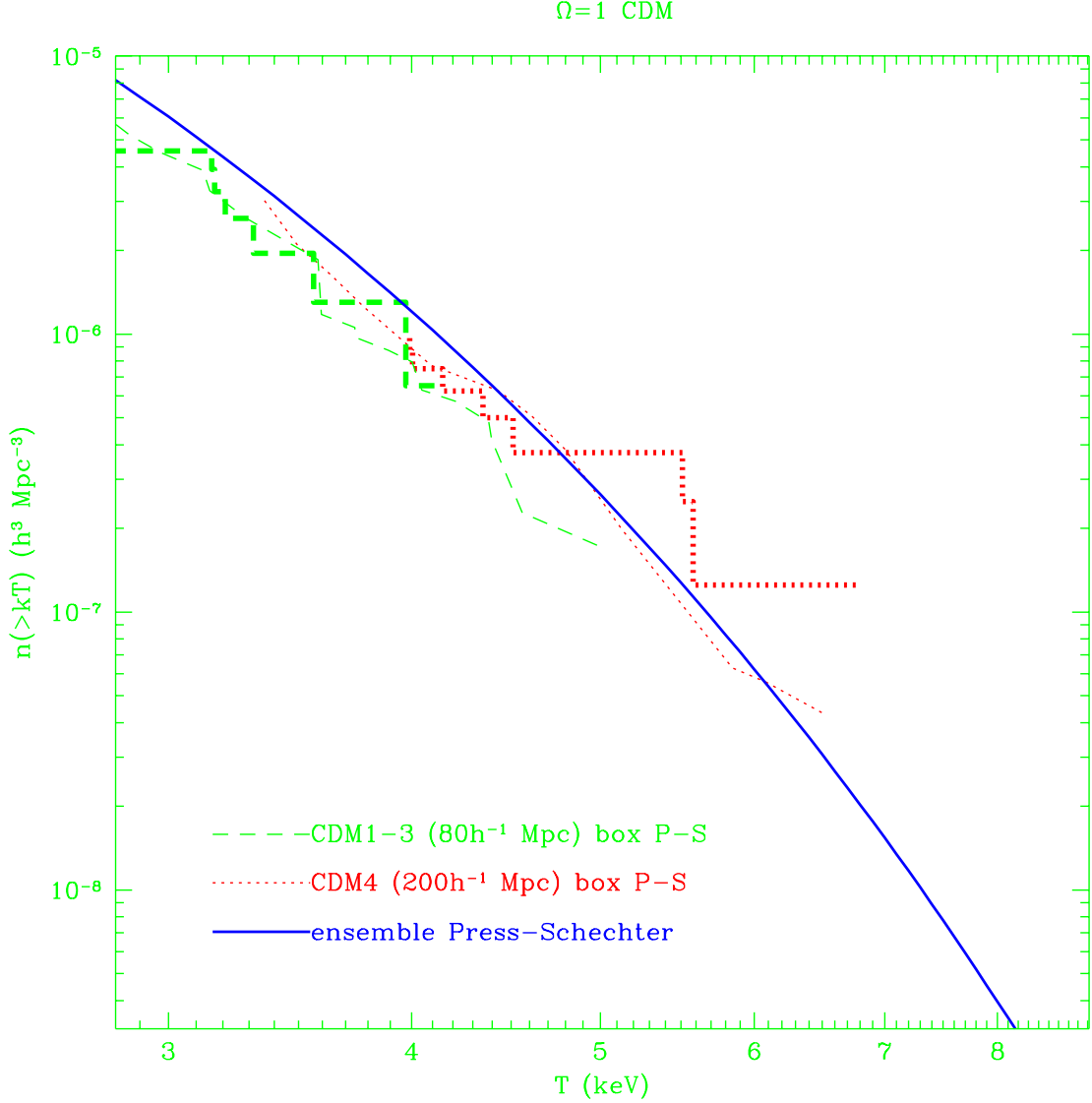}
\caption{The cumulative CDM $\sigma_8=0.5$ simulation temperature
function. 
The long diagonal curve is the Press-Schechter cluster abundance with
the best fit $T_8=4.9$keV.   $T_8$ is the temperature of a cluster
formed from the collapse of an $8h^{-1}$ Mpc radius sphere.
The  stepped lines were obtained from four CDM simulations.  The upper
left steps are from the three simulations CDM1-3 with
box size $80h^{-1}$ Mpc, and lower right steps are from CDM4 simulated
in a $200h^{-1}$ Mpc box.
The dashed diagonal lines are the Press-Schechter cluster abundance obtained
obtained with the 
actual simulation realizations.  For the smaller boxes, a loss of
power due to the absence of long waves causes the long dashed line to
be lower than the solid ensemble average.
}
\label{fig:cdmtemp}
\end{figure}

To solve for $T_8$ we identified clusters in each $\Omega_0=1$
simulation, resulting in the numerical cumulative distribution
$n(>T)$ shown as the  step lines in Figure \ref{fig:cdmtemp}.  We then
applied (\ref{eqn:asymptsimp}) to the linearly 
extrapolated density field of the simulation's initial conditions
$\delta \rho(z=0)=\delta \rho(z_i)\times (1+z_i)$.  This density field
is then smoothed on tophat spheres of varying radii, from which we
obtain the predicted Press-Schechter abundance of the collapsed mass
fraction shown as the dashed diagonal lines in Figure
\ref{fig:cdmtemp}.  We 
then solve for $T_8$ such that (\ref{eqn:asymptsimp}) 
agrees with the predicted abundance in the simulation volume for the 5
hottest clusters.  This approach compensates for the finite box size
effect which results in a loss of power on large scales (Pen 1997b).
The infinite volume average is the solid diagonal curve in Figure
\ref{fig:cdmtemp}.  We see that the smaller boxes systematically
underestimate the cluster abundance due to the suppression of
$\sigma_8$ from the truncation of the power spectrum in the finite box.

The best resulting fit is
\begin{equation}
T_8=4.9\pm .2 {\rm keV}.
\end{equation}
This value should be compared to $T_8=5.5$ keV for the perfect
isothermal sphere model (ECF).  We would expect the ECF model to
overestimate $T_8$ for several reasons.  The gas is almost certainly
not in perfect hydrostatic equilibrium, which will lower the
termperature.  A scatter in the mass-temperature relation also creates
more hot clusters due to Malmquist bias.  Cluster profiles may have
departures from isothermality, with slight temperature gradients throughout
the cluster.  The X-ray emission weighted temperature can be slightly
different from the mean mass weighted virial temperature.
ECF provide for a fudge factor
$\beta$ which describes the ratio between the temperature they used
and the actual statistical temperature of clusters.  Note that the
mean mass-temperature relation for an ensemble of clusters is not
sufficient to substitute correctly into the Press-Schechter relation
(\ref{eqn:asymptsimp}).  Since clusters are very rare, the mass
function is steep, and any scatter in the mass temperature relation
will introduce a bias in the abundance.  What we really want is to
normalize (\ref{eqn:asymptsimp}) directly to the temperature function
measured in simulations.  One can rescale the ECF result by choosing
$\beta=1.1$ in their temperature conversions.  Concurrent work by
Bryan and Norman (1997) produced results very similar to this study.
They defined a parameter $f_T$ as the ratio of the actual temperature
of a cluster compared to that obtained from the tophat
model with some assumed radial density profile.  In our notation, a
rough correspondence would be 
$f_T=T_8/5.5$, for which we would obtain $f_T=0.89\pm 0.03$, while
they found values in the range $0.75 \lesssim f_T \lesssim 0.92$.

The error interval is the $1-\sigma$ standard deviation comparing the
different simulations and represents departures from a deterministic
P-S theory.  To first order, we have compensated for numerical
cosmic variance, and also the loss of large scale power from the
finite box size.  The simulations only have a limited sampling volume,
and do not directly simulate the rarest, richest clusters.  By using
the P-S formalism, we can extrapolate the simulation normalization to
larger volumes by assuming the mass temperature relation scales as one
would expect from virial equilibrium.  Each simulated model still had
at least one cluster above the pivot temperature of 5 keV.  In this
analysis, P-S allows us to reduce the error bars by simultaneously
fitting a range in cluster temperatures.

The first non-gravitational effect which needs to be incorporated is
the effect of heat injection from stars.  Direct observations of the
iron line emission suggests that the intra-cluster medium (ICM) is not
pristine, and has passed through an earlier generation of stars.  This
may have raised the initial entropy of the gas.  The present day
metallicity of the ICM is near $1/3$ solar, from which we may infer up
to 1 keV of energy per nucleon to have been injected (Loewenstein and
Mushotzky 1996).  While it is not known when this might have happened,
clusters at redshift 0.3-1 appear to have similar metalicities as
nearby clusters (Mushotzky and Loewenstein 1997, Hattori \etal\ 1997).
Early enrichment would have a smaller effect since the adiabatic
expansion of the universe cools the preheated gas.  In model PREHEAT
we inject gas consistent with the observed lack of evolution to $z=1$.
The most extreme model postulates 1 keV of energy injection at $z=1$.
We simulated such a model by evolving a simulation to $z=1$, raising
the thermal energy everywhere by 1 keV, and continuing the evolution
to $z=0$.  We then measured the present cluster temperature function.
We find that clusters are slightly hotter with $T_8=5.3$ keV.  We
should consider this an upper limit on the plausible effect of
heating.  Because the gas remains in hydrostatic equilibrium with the
dark matter potential, the injection of 1 keV only raised the mean
temperatures by 0.4 keV.  For our choice of flat universe parameters,
this lowers the normalization to $\sigma_8=0.50$ relative to the
adiabatic value of 0.53 (see below).  Future work will
also address the effects of cooling (Cen \etal\ 1998).  Fabian (1994)
estimates that up to 20\% of the X-ray luminosity of a cluster may
arise from a cooling flow, which might affect the emission weighted
temperature by a similar amount as the heating.  The latter depends
sensitively on the smallest scale inhomogeneities, and poses a larger
computational challenge.  For now we will assume temperature errors to be
symmetric.

\begin{figure}
\plotone{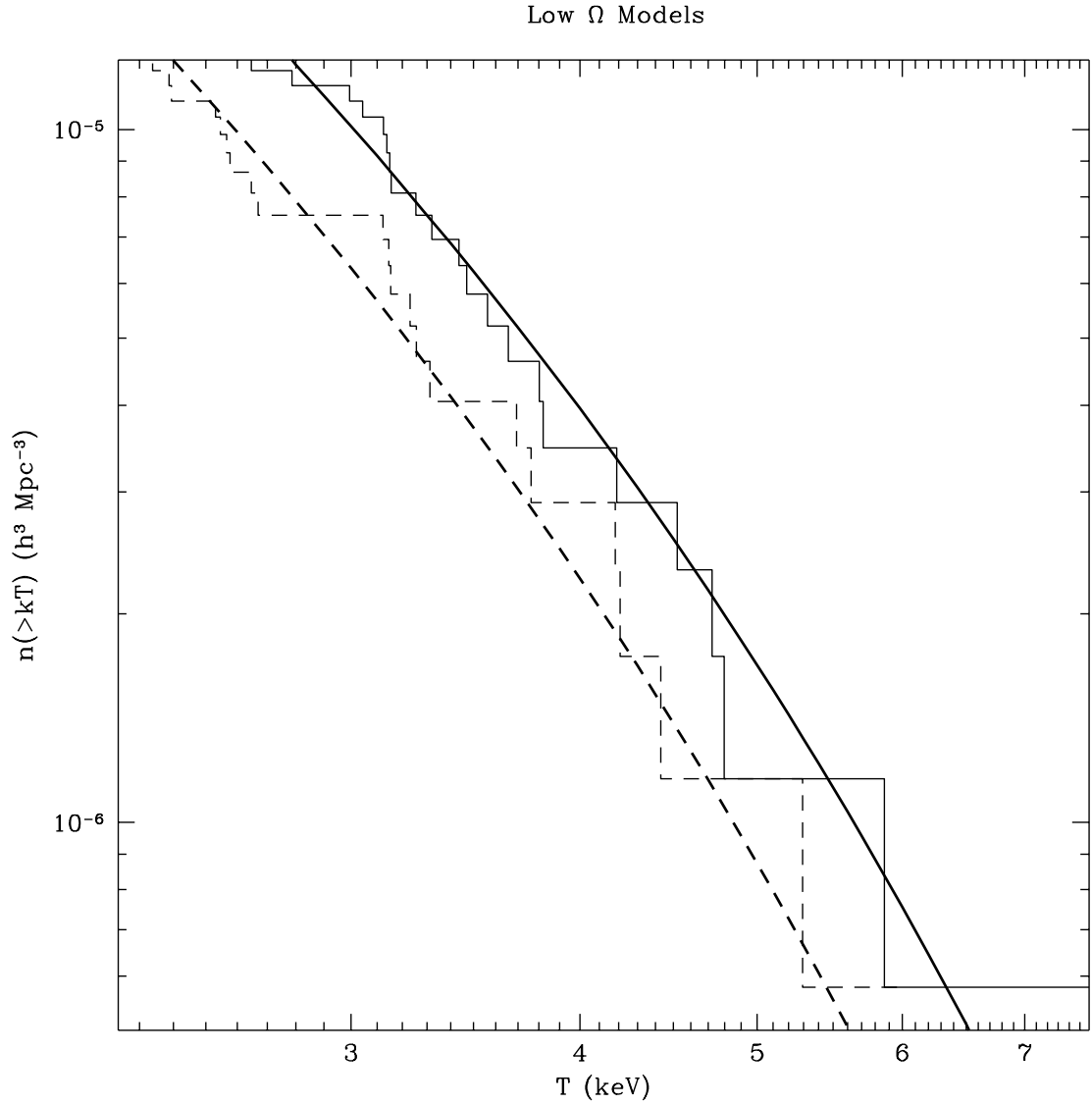}
\caption{The simulation fits to hyperbolic and cosmological constant
dominated universes.  Both simulations had identical random seeds and
$\Omega_m=0.37$ and $\sigma_8=1$.  The dashed abundance is for the
cosmological constant 
model.  The solid line is the hyperbolic model.}
\label{fig:olcdm}
\end{figure}

Once we have solved for $T_8$ using the $\Omega_0=1$ simulations, we
proceed to repeat the procedure for the $\Lambda$ and hyperbolic
cosmologies.  By using the same random seed, we expect the differences
between the models to be modeled more accurately than each model
individually.  The $\Omega$ dependence is incorporated into to
mass-temperature relation as
\begin{equation}
T_8=4.9\pm .2
\Omega_0^{2/3}\Omega(z)^\gamma {\rm keV}.
\label{eqn:mtr}
\end{equation}
The $2/3$ scaling accounts for
the smaller virial mass enclosed in the $8h^{-1}$ Mpc spheres as the
density is lowered.  $\gamma$ will be used to parameterize remaining
corrections, such as the change in formation redshift, change in
virial radius and collapse density.  
Solving for $\gamma$ from the $\Lambda$ simulation
yields $\gamma=0.283$ for a $\Lambda$ universe and $\gamma=0.133$ in the
hyperbolic scenario.  The temperature function fits for the simulated
low density parameters are shown in Figure \ref{fig:olcdm}.

\begin{figure}
\plotone{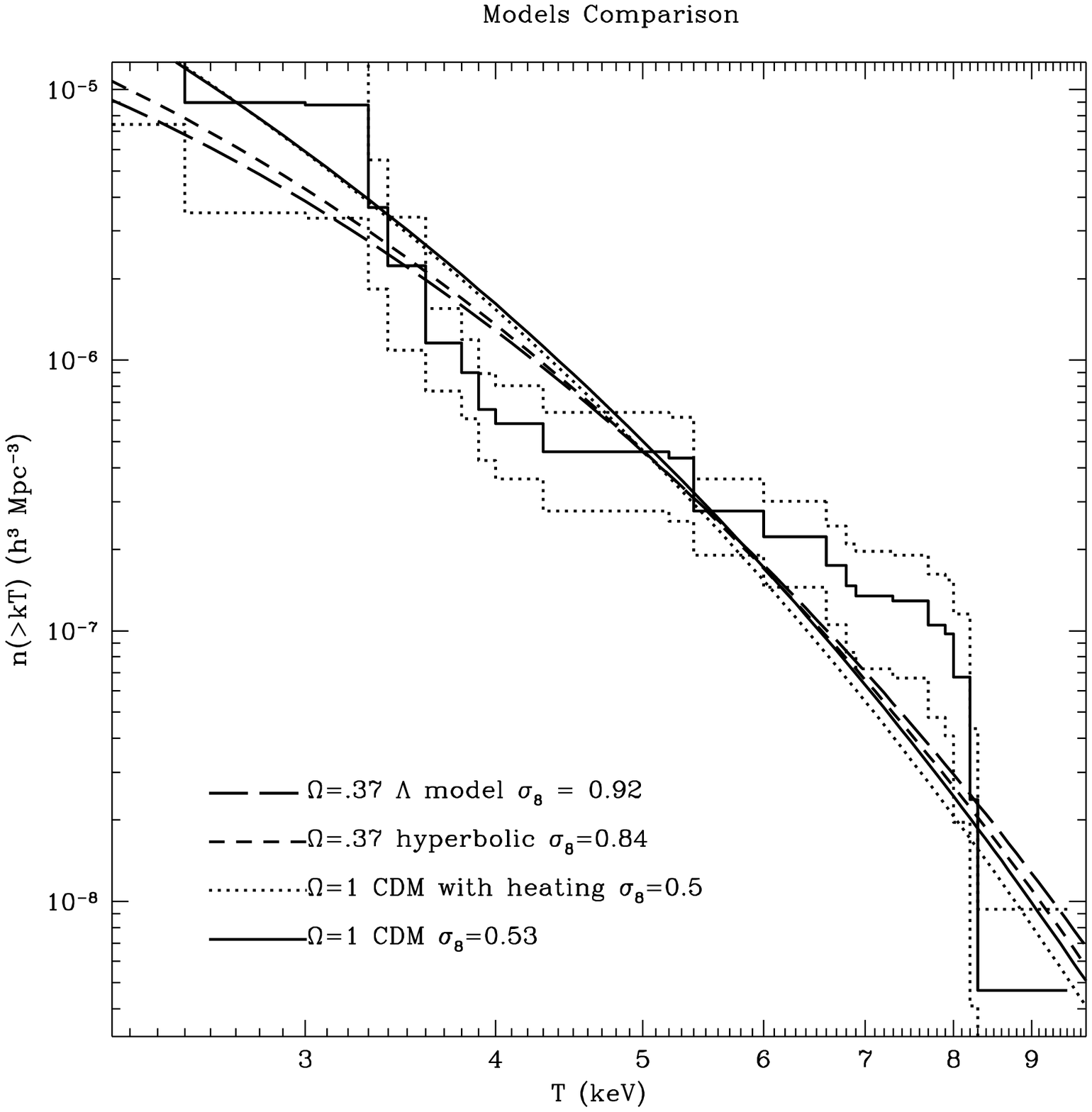}
\caption{Comparison of the best fit normalizations to the Henry and
Arnaud cluster sample.}
\label{fig:fits}
\end{figure}

We now have all the required relations to solve for $\sigma_8$ given
the cluster abundance.  For a given $\Omega_0$, we have the left hand
side of Equation (\ref{eqn:asymptsimp}) from HA at any given
temperature, for which we follow ECF and use kT = 5 keV.  Equations
(\ref{eqn:mtrz}) and (\ref{eqn:mtr}) allow us to convert the
temperature into $m$.  The only remaining unknown variable is
$\sigma_8$ which determines $\mu$ and $v$.  We then obtain the result
$\sigma_8 =0.53\pm 0.05
\Omega_0^{-0.53}$ in the $\Lambda$ model, and $\sigma_8=0.53\pm 0.05
\Omega_0^{-0.45}$ in the hyperbolic universe.  The error bars are
obtained by linearly adding the uncertainty in the mass-temperature
relation to the effects of supernova heat injection.  Figure
\ref{fig:fits} summarizes the results from of the numerical
normalizations.  In the hyperbolic model, matter domination ends at a
relatively higher redshift $z \sim (1/\Omega)-1$ than in a cosmological
constant model.  Cluster of the same mass will thus have a smaller
radius in the hyperbolic case, and thus a higher virial temperature.
This accounts for the smaller value of $\gamma$ and $\sigma_8$ in
hyperbolic models.  Our modeling of the temperature function
(\ref{eqn:mtr}) implicitly accounts for all these effects since it is
normalized by simulations.


\section{Cosmological Implications}
\label{sec:cosmo}

In the context of structure formation, the simplest adiabatic models,
which could arise for example from inflation, have only a
single free parameter to normalize the spectrum of fluctuations for a
HZP spectrum.  For a COBE normalized flat universe,
this implies $\sigma_8=1.2$, which is at great odds with the observed
cluster abundance.  If we wish to retain a flat universe, one can can
try to change one of several parameters.  The first would be the
baryon fraction $\Omega_b$ (White \etal\ 1995) which suppresses
fluctuations due to acoustic oscillations and Silk damping.  COBE
normalized CDM agrees with the cluster abundance for $\Omega_b=0.45$,
which requires a dramatic revision of Big-Bang nucleosynthesis (Walker
\etal\ 1991).  A different parameter which can lower the COBE
normalized value of $\sigma_8$ is a tilted power spectrum,
i.e. deviations from HZP.  Leaving only this one
parameter free, we find a satisfactory fit for $n=0.63$ when we also
allow for tensor modes.  This violates the limits on the slope allowed
by the 4 year COBE data (Wright \etal\ 1996).  A combination of these
two parameters violates each of these constraints more weakly, and one
could envision combinations such as $n=.7,\ \Omega_b=.15$, which would
also be consistent with the cluster gas fractions (White \etal\ 1993,
White and Fabian 1995).  One can also take more radical departures,
and lower the Hubble constant to $h\sim 0.3$ (Bartlett \etal 1995).
We conclude that no single parameter modification of $\Omega=1$
inflationary cosmology is even marginally consistent with observations.

For low $\Omega_0$ models, the opposite problem arises.  COBE
normalized fluctuations result in too low values of $\sigma_8$.  For
the hyperbolic universe with parameters $\Omega_0=0.37$, $h=0.7$, we
obtain $\sigma_8=0.57$, significantly lower than the $0.83$ suggested
by the cluster abundance.  To raise it to match the cluster abundance,
one can either raise $\Omega_0$ to 0.43, raise the Hubble constant to
$h=1.1$, or introduce a tilt $n=1.17$, or any combination thereof.
The cosmological constant models also have their share of free
parameters.  By lowering $\Omega_0$, we lower the normalization, but it
increases with larger Hubble constant $h$.  For the choices under
discussion $\Omega_0=0.37$, $h=0.7$, we have COBE normalized $\sigma_8
= 1.0$, just slightly higher than the cluster abundance suggests.
This can easily be addressed by lowering $h$ to 0.63, or using the
slight tilt $n=.95$ as suggested by Ostriker and Steinhardt (1995).

A third alternative is to consider non-adiabatic initial conditions,
for example from topological defects (Pen \etal 1997).  In
these models the P-S abundance must be modified to account for the
non-Gaussianity (Chiu \etal 1997).  These models have not been studied
directly with hydrodynamical simulations of the cluster temperature
function, but the preliminary results indicate a cluster abundance
consistent with a COBE normalized HZP spectrum.  We note that these
models have problems with other observations, including the galaxy
power spectrum and small scale microwave background anisotropies.

The HA sample has the great advantage that it has well established
completeness criteria, which allows us to accurately measure cluster
abundances.  Many more clusters have measured temperatures, and one can
ask how those might affect our estimates.  Most importantly, we at not
limited by Poissonian statistics but rather by systematic errors.  It
has been proposed that the evolution of the temperature function is a
strong test of cosmologies.  At fixed temperature, the difference in
cluster abundance at $z=0.5$ is over an order of magnitude between flat
and hyperbolic models (ECF).  Again, we are not limited by statistics.
Instead, we must ask what the expected difference in cluster
temperature is at fixed abundance.  Presently, the hottest cluster in a
$300 h^{-1}$ Mpc radius is about 8 keV (see Figure \ref{fig:hadata}).
At a redshift $z=0.3$, the hottest cluster in the same sample volume
would be essentially the same temperature in a hyperbolic universe, and
about 6 keV in a flat universe.  One must make sure that systematic
temperatures measurement errors are less than 25\% at these redshifts.
Since the metallicities have not evolved significantly, it appears that
heating will not contribute significantly to the cluster temperature
evolution.  The presence of hot clusters such as A2163 and MG2016
(Hattori \etal\ 1997) may well pose problems for flat cosmologies (Pen,
David and Tucker 1997).  Recently, Carlberg \etal (1997) studied the
abundance of galaxy clusters at intermediate redshift using galaxy
velocity dispersions (see also Fan \etal 1997).  The difficulty here is
converting local cluster temperatures into velocity dispersions (Bird
\etal 1995, Bahcall and Lubin 1995) since the expected change is
small.  Any scatter in the velocity-temperature relation will introduce
systematic biases which need to be understood.  A systematic
overestimate of 10\% in the line-of-sight velocity dispersion
$\sigma_v$ in the CNOC sample relative of the converted $\sigma_v$ from
the HA temperature data is sufficient to offset the decrease in the
cluster abundance predicted in an $\Omega=1$ cosmological model.  It is
essential in these comparisons that homogeneous samples are used at
both low and high redshift which are checked directly against
simulations.

Clusters of galaxies provide, coincidentally, a similar constraint on
$\Omega_0$ as velocity field measurements.  Measurements of peculiar
velocities constrain $\beta_v \equiv \Omega_0^{.6}/b$, where the bias
$b$ is the ratio of fluctuation in galaxies relative to the dark
matter (Strauss and Willick 1995).  Typical values for $\beta_v$ are
in the range $0.3-0.8$.  Cluster abundances from the temperature
function constrain a very similar function $\sigma_8 \Omega_0^{\sim 0.6}
\sim 0.5$, where $\sigma_8 = 1/b$ for optical galaxies.  While
velocity fields measure fluctuations in the linear regime, the cluster
abundances are determined for highly non-linear bound objects.  It is
reassuring that the values obtained from the two very different
methods are consistent with each other.  The downside is that we
cannot determine $\Omega_0$ and $b$ independently using present day
measurements alone.

\section{Conclusions}

The most robust cosmological constraints using clusters of galaxies
come from the cluster temperature function, which depends primarily on
the gravitational potential wells of the dark matter.  We used new gas
dynamic simulations to test the N-body and Press-Schechter estimates
by ECF and VL.  We found good general agreement in the normalization
to the observed cluster temperature function with $\sigma_8 = 0.53\pm
0.05 \Omega_0^{-0.45}$ for a hyperbolic universe, and
$\sigma_8=0.53\pm 0.05 \Omega_0^{-0.53}$ for a spatially flat universe
with a cosmological constant.  This result is only weakly sensitive to
models of the thermal history of the intra-cluster medium, which we
have modeled with preheating.  We have presented improved
Press-Schechter fits to predict the mass-temperature-relation scalings
(\ref{eqn:mtr}) more accurately for a range in values of $\Omega_0$.
Applications to high redshift X-ray clusters are in progress (Pen,
David and Tucker 1997).

COBE normalization with a Harrison-Zeldovich spectrum over-predicts
cluster abundances by many orders of magnitude, but this can be
addressed if the spectrum is strongly tilted and a large baryon
fraction is invoked.  The $\Omega_0=0.35$ hyperbolic or cosmological
constant models lie closer to observations on all measures, including
the age of the universe, the slope of the temperature function, the
gas fraction and the slope of the galaxy power spectrum just to
mention a few.  While no single measurement is at a very high
significance, the combination does appear to carry a heavy vote.
Lensing statistics (Kochanek 1996) and deceleration parameter
measurements (Perlmutter \etal 1997, Pen 1997c) would favor a
hyperbolic universe over one with a cosmological constant, and
alternative scenarios (for example the string dominated model, Spergel
and Pen 1996) are also viable.  COBE normalized topological defect
models fare reasonably well on the cluster abundance (Chiu \etal
1997).

\begin{center}
Acknowledgements
\end{center}

I wish to thank Professor J.P. Ostriker for his continuous support
during this work, and Professor M.A. Strauss for significant feedback.
I thank Mike Balogh for a careful checking of the equations.  Financial
support was provided by Princeton University through the Porter Ogden
Jacobus Fellowship, NSF grant ASC93-18185, the Harvard-Smithsonian
Center for Astrophysics, and the Harvard Society of Fellows.  Computing
time was provided by the National Center for Supercomputing
Applications and the Pittsburgh Supercomputer Center.  Stimulating
discussions with Uros Seljak and Avi Loeb were helpful during the
course of this work.

\begin{thebibliography}{}
\bibitem{} Arfken, G 1985, ``Mathematical Methods for Physicists''.
\bibitem{}  Bahcall, N. 1996, in ``Astrophysical Quantities'' by Steve
Allen, 2nd Ed.
\bibitem{} Bahcall, N., and Lubin, L.M. 1994, \apj, 426, 513.
\bibitem{}  Bardeen, J.M., Bond, J.R., Kaiser, N., \& Szalay, A.S. 1986
\apj, 304, 15.
\bibitem{}  Bartlett, J.G., Blanchard, A., Silk, J. and Turner, M. S.
1995, Science, 267. 980.
\bibitem{} Bryan, G.L. and Norman M.L. 1997, astro-ph/9710187.
\bibitem{} Carlberg, R.G., Morris, S.L., Yee, H.K.C., Ellingson,
E. 1997, \apj, 479, L19.
\bibitem{} Cen, R., Bode, P., Ostriker, J.P. and Pen, U. 1998, in
preparation. 
\bibitem{} Chiu, W.A., Ostriker, J.P. and Strauss, M.A. 1997,
astro-ph/9708250. 
\bibitem{} Edge, A.C., Steward, G.C., Fabian, A.C., Arnaud, K.A. 1990,
\mnras, 245, 559.
\bibitem{} Eke, V.R., Cole, S., Frenk, C.S. 1996, MNRAS 282, 263.
\bibitem{} Fabian, A.C. 1994, \araa, 32, 277.
\bibitem{} Fan, X., Bahcall, N.A., and Cen, R. 1997, \apjl, 490, L123.
\bibitem{} Frenk, C.S., White, S.D.M. and others 1997, in preparation.
\bibitem{} Gelb, J.M. and Bertschinger, E. 1994, \apj 436, 491.
\bibitem{} Hattori, M., Ikebe, Y., Asoaka, I., Takeshima, T.,
B\"oringer, H., Mihara, T., Neumann, D.M., Schindler, S., Tsuru, T.,
Tamura, T. 1997, ``Discovery of the most distant X-ray galaxy cluster
in the direction of the gravitational lens system MG2016+112'',
preprint. 
\bibitem{} Henry, J.P. and Arnaud, K.A 1991, \apj, 372, 410.
\bibitem{} Holtzman, J.A. 1989, \apjs, 71, 1.
\bibitem{} Kang, H., Cen, R., Ostriker, J.P., Ryu, D. 1994, \apj, 428,
1.
\bibitem{} Kochanek, C.S. 1996, \apj 466, 638.
\bibitem{}  Loewenstein, M., Mushotzky, R. F. 1996, \apj,466,695.
\bibitem{} Mushotzky, R. F. and Loewenstein, M. 1997, \apj, 481, 63.
\bibitem{} Ostriker, J.P., and Steinhardt, P.J. 1995, Nature, 377, 600.
\bibitem{} Pen, U. 1995, \apjs, 100, 269.
\bibitem{} Pen, U. 1997a, ``A High-resolution  Adaptive Moving Mesh
Hydrodynamic Algorithm'', \apjs, in press, astro-ph/9704258.
\bibitem{} Pen, U. 1997b, \apjl , 490, L127.
\bibitem{} Pen, U., Seljak, U. and Turok, N. 1997, \prl, 79, 1611.
\bibitem{} Pen, U. 1997c, New Astronomy, 2, 309.
\bibitem{}   Pen, U., and Spergel, D.N. 1995, \prd 51, 4099.
\bibitem{} Pen, U., David, L. and Tucker, W.H. 1997, in preparation.
\bibitem[Perlmutter \etal 1996b]{perl96b}  Perlmutter, S.,
Gabi, S., Goldhaber, G.,
Groom, D., Hook, I., Kim, A., Kim, M., Lee, J., Pennypacker, C.,
Small, I., Goobar, A., Ellis, R., McMahon, R., Boyle, B.,
Bunclark, P., Carter, D., Irwin, M., Glazebrook, J., Newberg, H.,
Filippenko, A.V., 
Matheson, T., Dopita, M. and Couch, W.  1997, \apj, 483, 565.
\bibitem{} Press, W.H. and Schechter, P. 1974, \apj, 187, 425.
\bibitem{} Spergel, D.N. and Pen, U. 1996, ``a String Dominated
Universe'', astro-ph 9611198, \apjl, in press.
\bibitem{} Strauss, M.A. and Willick, J.A. 1995, Physics Reports, 261,
271.
\bibitem{} Viana, P.T.P. and Liddle, A.R. 1996, \mnras, 281, 323.
\bibitem{}  Walker, T. P., Steigman, G., Kang, H., Schramm, D. M.,
Olive, K. A. 1991,  \apj, 376, 51
\bibitem{}  White, D.A., and Fabian, A.C. 1995, \mnras, 273, 72.
\bibitem{} White, M., Scott, D., Silk, J., Davis, M., 1995, \mnras,
276, 69.
\bibitem{} White, S.D.M., Navarro, J.F., Evrard, A.E. \& Frenk,
C.S. 1993, Nature, 366, 429
\bibitem{} Wright, E. L., Bennett, C. L., Gorski, K., Hinshaw, G.,
Smoot, G. F. 1996, \apj, 464, 21.
\bibitem{}  Xin, Z. and Jin, S. 1995, Comm. in Pure and Applied Math,
48, 235. 
\end{thebibliography}
\end{document}